\documentclass[
aps,
reprint,
prl,
superscriptaddress,
amsmath,
amssymb,
floatfix,
longbibliography,
]{revtex4-2}

\usepackage{bm}
\usepackage{graphicx}
\usepackage{latexsym}
\usepackage{array}
\usepackage{amsfonts}
\usepackage[free-standing-units]{siunitx}

\usepackage{mathtools}  
\usepackage{makecell}   
\usepackage{hhline}     
\usepackage{amsthm}
\usepackage{amscd}

\newcommand{\CH}[1]{{\color{red}[#1]}}
\newcommand{\AM}[1]{{\color{blue}[#1]}}
\newcommand{\SB}[1]{{\color{magenta}#1}}

\usepackage{xcolor}
\renewcommand\vec{\boldsymbol}

\definecolor{orange}{rgb}{1,0.5,0}
\definecolor{goodGreen's}{rgb}{0.1,0.5,0}
\definecolor{goodred}{rgb}{0.7,0,0}
\usepackage[colorlinks,urlcolor=goodGreen's,citecolor=blue,linkcolor=goodred]{hyperref}

\newcommand{\orcid}[1]{\href{https://orcid.org/#1}{\includegraphics[width=8pt]{orcid.png}}}

\usepackage{verbatim}
\usepackage{hyperref}
\usepackage[normalem]{ulem}
\graphicspath{{images/}}

\begin{document}

\title{Phase-induced switching of ferromagnetic insulators in  Josephson spin valves}

\author{A. A. Mazanik  }
\email{andrei.mazanik@csic.es}
\affiliation{Centro de Física de Materiales (CFM-MPC) Centro Mixto CSIC-UPV/EHU,
E-20018 Donostia-San Sebastián, Spain}

\author{C.-H. Huang    }
\email{chenhow.huang@gmail.com}
\thanks{ These authors contributed equally to this work.  }
\affiliation{Department of Physics and Nanoscience Center, University of Jyväskylä, P.O. Box 35 (YFL), FI-40014 University of Jyväskylä, Finland}

\author{Miguel A. Cazalilla} 
\affiliation{Donostia International Physics Center (DIPC), 20018 Donostia-San Sebastian, Spain}
\affiliation{Ikerbasque, Basque Foundation for Science, 48013 Bilbao, Spain}

\author{F. S. Bergeret}
\affiliation{Centro de Física de Materiales (CFM-MPC) Centro Mixto CSIC-UPV/EHU, E-20018 Donostia-San Sebastián, Spain}
\affiliation{Donostia International Physics Center (DIPC), 20018 Donostia-San Sebastian, Spain}

\date{\today}

\begin{abstract}
We study the Josephson effect in junctions composed of two ferromagnetic insulator/diffusive superconductor bilayers  separated by an insulating barrier. By computing the free energy of the system, we identify two distinct contributions: (i) The work performed by a current source to create a supercurrent through the junction, and (ii) an antiferromagnetic coupling between ferromagnetic insulators, mediated by the superconducting condensate across the insulating barrier. The competition between these contributions allows for switching between parallel and antiparallel configurations of the magnetizations of the ferromagnetic insulators. We explicitly show that the switching occurs at finite temperatures and for superconducting phase differences satisfying $\pi/2 < \phi < 3\pi/2$. Importantly, this effect can be realized in ferromagnetic insulators with sufficiently large easy-plane anisotropy energy. Using realistic junction parameters, we demonstrate that the switching can be controlled by phase bias and triggered by half–flux-quantum voltage pulses or external magnetic field pulses on the microsecond timescale. These results provide a  route towards controllable Josephson-based superconducting memory devices based on  EuS/Al heterostructures.
\end{abstract}

\maketitle

\def\thefootnote{*}\footnotetext{These authors contributed equally to this work.}

When two superconductors are brought into contact through a thin insulating layer, Cooper pairs can undergo quantum tunneling between them. This is the essence of the Josephson effect~\cite{barone1982physics}, which is described by the following mathematical expression:
%
\begin{equation} \label{eq:Usual_Josephson_energy}
    E_{JJ}(\phi)   = \frac{I_{c}}{2e} \left( 1 - \cos \phi \right),
\end{equation}
relating the phase difference between the superconductors, $\phi$, with the Josephson energy, $E_{JJ}$.
Here, 
$I_{c}$ is the critical current of the junction, and $e$ is the free electron charge. 
The Josephson energy is the work done by an external  source to create a supercurrent,
\begin{equation} \label{eq:CPR0}
    I_S(\phi) =2 e  \frac{\partial E_{JJ}}{\partial \phi} = I_c \sin\phi,
\end{equation} 
through the junction starting from a state without any supercurrent.  The relation~\eqref{eq:Usual_Josephson_energy} and the current-phase relation (CPR), Eq.~\eqref{eq:CPR0}, have been thoroughly investigated in many types of superconducting systems. These studies encompass 
conventional and unconventional superconductors, as well as junctions containing normal metals, semiconductors, magnetic materials, and point contacts. This body of work has been thoroughly reviewed in books  
\cite{barone1982physics,tafuri2019fundamentals} and  review articles \cite{Golubob2004,Bobkova_2022,Shukrinov_2022,Birge24}.

In terms of applications to modern information technologies,
Josephson junctions (JJs) are platforms that hold great promise. 
For instance, the possibility of implementing JJ-based memory and synaptic connections has attracted a great deal of attention in recent years~\cite{Soloviev2017,Schneider18,schneider2020synaptic,Schneider22}. A strategy to achieve this goal is to engineer JJs whose critical current can be tuned by an external control parameter. Such a tunability can be realized in Josephson spin valves, i.e. hybrid structures consisting of superconductors and, at least, two ferromagnetic layers, metallic or insulating~\cite{banerjee2014reversible,Krasnov14,gingrich2016controllable,Birge18,Naaman18,Giazotto18,Madden_2019,Baek15} (see also   Ref.~\cite{Birge24}, for a recent review). In these devices, the amplitude and sign of the critical current is determined by the relative orientation of the magnetizations of the ferromagnetic elements, which is controlled by an external magnetic field or dissipative spin currents.


Beyond the foregoing control schemes, several theoretical 
works~\cite{WaintalBrouwer02,Sauls08,Linder11,Linder12,Linder14,Halterman15,Halterman_2016,Shomali_2011,Shomali_2020} have predicted that, in fully-metallic Josephson spin valves, non-collinear magnetizations can create equilibrium spin currents, tunable by the phase difference $\phi$. Such spin currents allow one to control the relative orientation of layer magnetizations  using $\phi$. However, no experimental realization of phase-induced switching has been reported to date. We deem the main obstacle to the realization of this prediction the requirement of strong ferromagnetic elements~\cite{Shomali_2011,Shomali_2020}, which  substantially suppress superconductivity. Moreover, strong ferromagnets and the interfaces between them and other metals are also an important source of strong spin relaxation \cite{Bass_2007}, further weakening the effect.

\begin{figure} [hbtp]
    \centering
    \includegraphics[width=0.7\linewidth]{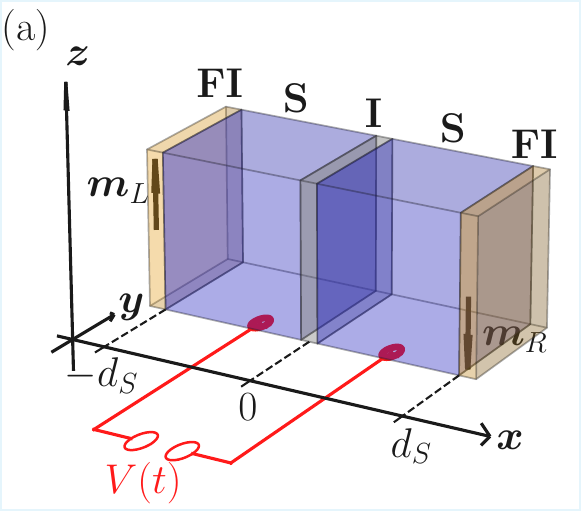}\\
    \includegraphics[width=0.45\linewidth]{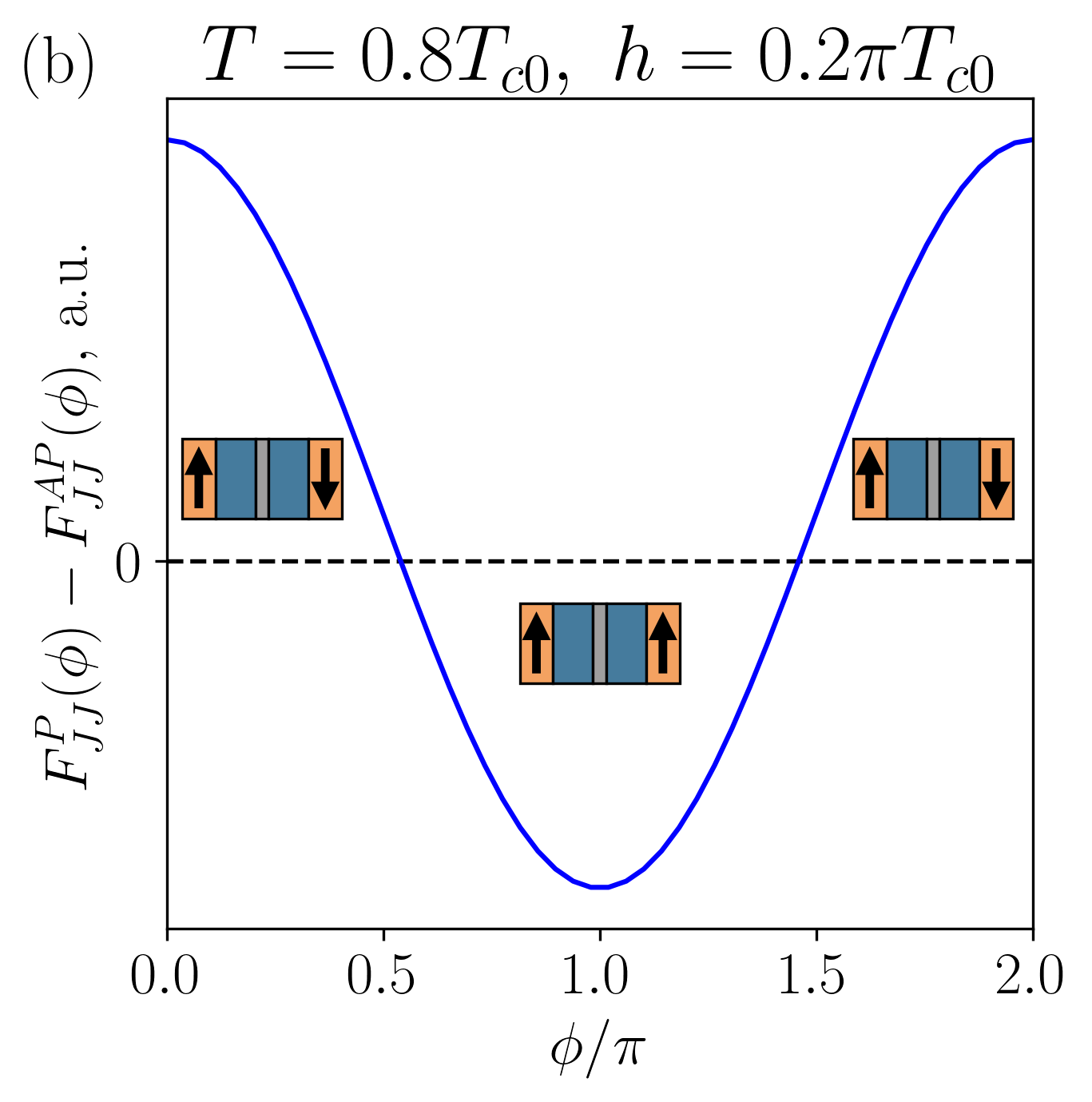}
    \includegraphics[width=0.45\linewidth]{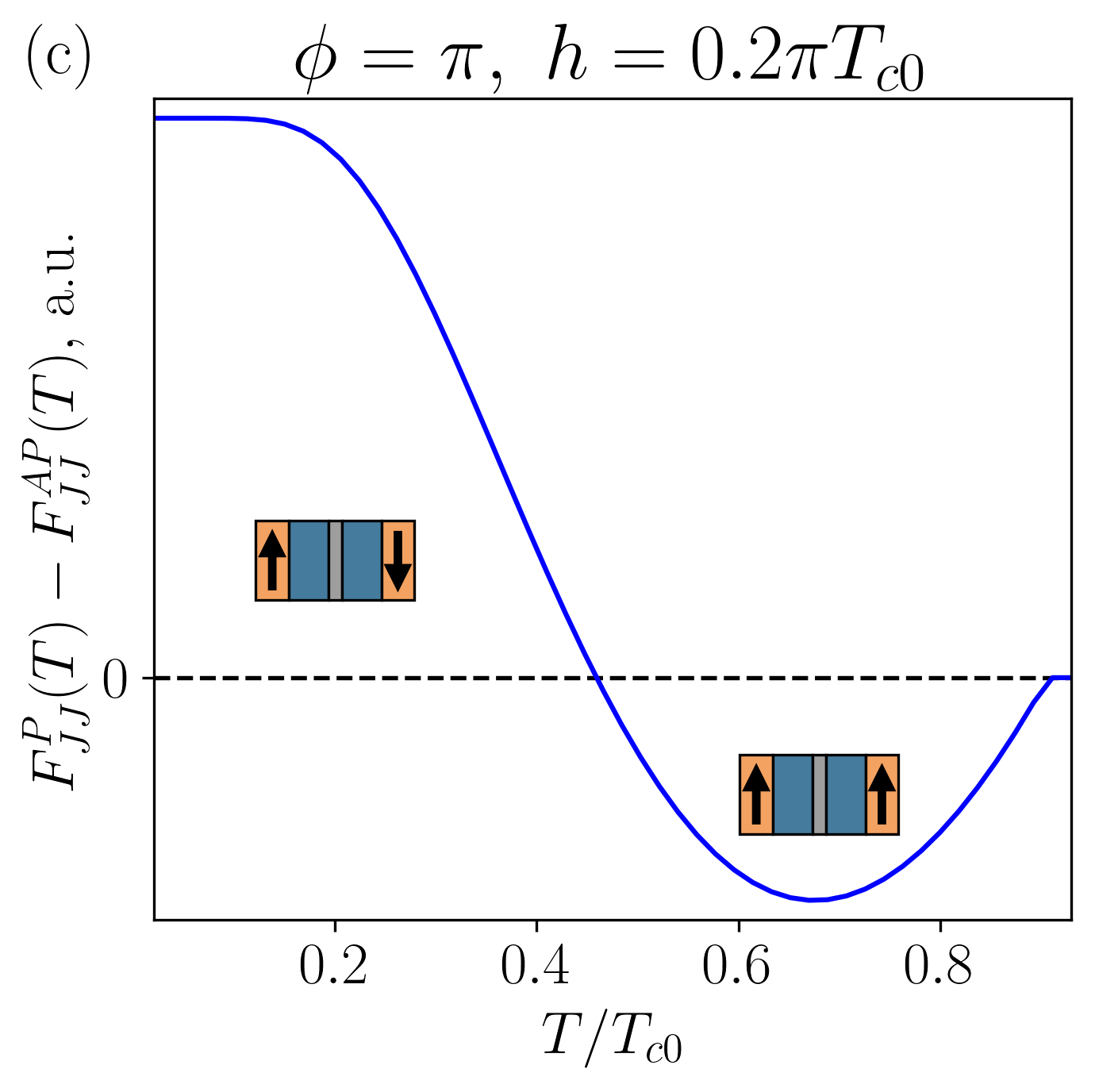}
    \caption{(a): Sketch of the Josephson junction considered in this work. It consists of five layers, FI/S/I/S/FI. Each superconducting (S) layer has a width $d_S$ and each ferromagnetic insulator (FI) layer produces an exchange field $\vec{b}_{L,R} a\delta(x \pm d_S)$ in each superconductor. The $\vec{b}_{L,R}$ of the FIs are aligned along unit vectors $\vec{m}_{L,R}$. A (time-dependent) voltage bias $V(t)$ is applied to the superconducting elements  fixing the phase difference between them to $\phi(t) = \phi(-\infty)  + \int_{-\infty}^{t} 2eV(t')dt'$. (b,c): Sketch of the difference in tunneling energies between Josephson junctions with the FIs in the parallel (P) and antiparallel (AP) configurations, $F^{P}_{JJ} - F^{AP}_{JJ}$,  as a function of the phase difference (b) and of the system temperature (c). }
    \label{fig:setup}
\end{figure}

In this Letter, we investigate the phase-induced magnetization switching in a different system. Namely, we study the Josephson spin valve shown in Fig.~\ref{fig:setup}.  The system consists of two superconducting layers adjacent to two distinct ferromagnetic insulators (FIs). Due to the magnetic proximity effect, the superconductors are spin-split~\cite{moodera1988electron,hao1991thin,Strambini17,Giazotto18,Hijano21}. In addition, when separated by an insulating barrier (I), they form a Josephson junction. Using realistic parameters, we demonstrate that  phase-induced magnetization switching is possible in our junction, free from the drawbacks of all-metallic valves: the effective exchange field remains smaller than the superconducting gap, whereas spin relaxation is limited only by the superconductor itself, and it is typically weak. Moreover, the switching requires neither spin-orbit coupling~\cite{Guarcello20,Mazanik20,Bobkova_2022,Shukrinov_2022} nor external magnetic fields~\cite{banerjee2014reversible,Krasnov14,gingrich2016controllable,Birge18,Naaman18,Giazotto18,Madden_2019}, or dissipative spin currents~\cite{Baek15}. Importantly, our system can serve as a platform for cryogenic memory applications, as suggested in Ref.~\cite{Giazotto18}. However, unlike the dissipative system studied in~\cite{Giazotto18}, in which properties are governed by quasi-particle transport, our junction operates in a non-dissipative regime with switching induced by the superconducting phase difference.

\textbf{Model:} Each spin-split superconductor is realized as a ferromagnetic insulator/superconductor (FI/S) bilayer, resulting in a FI/S/I/S/FI structure for the Josephson junction. The superconductors have a width $d_S$ along the (super-)current direction. The magnetic proximity effect from the FIs produces  exchange fields $\vec{b}_{L,R}a \delta(x \pm d_S)$ in the adjacent superconductors. The local exchange fields $\vec{b}_{L}$ and $\vec{b}_R$ are assumed to be equal in magnitude, $\vert \vec{b}_L \vert = \vert \vec{b}_R\vert = b$, and aligned along the  unit vectors  $\vec{m}_{L}$ and $\vec{m}_R$, respectively. The Hamiltonian of the system reads
\begin{equation} \label{eq:H_total}
    \hat{H} = \hat{H}_L + \hat{H}_R + \hat{H}_T,
\end{equation}
where $\hat{H}_{L}$ and $\hat{H}_R$ are the BCS Hamiltonians of the left and right leads, respectively. We consider that the superconducting leads are in the dirty regime, characterized by a diffusion coefficient $D$, and that their width $d_S$ is much smaller than the superconducting coherence length $\xi = \sqrt{\frac{D}{2\pi T_{c0}}}$, where $T_{c0}$ is the critical temperature of the leads in the absence of FIs.  In this regime, both superconductors have a homogeneous spin-split spectrum with a spin-splitting field $h = ba/d_S$ \cite{Strambini17,Giazotto18,Hijano21}. In addition, we set the superconducting order parameters equal to each other, $\Delta_L = \Delta_R = \Delta$. The tunneling Hamiltonian, $\hat{H}_T$, is 
\begin{equation} \label{eq:tunneling_coupling}
    \hat{H}_T =  \sum_{\vec{p}\vec{q}}  \left[  \Psi^\dagger_L (\vec{p}) \check{t}_{\vec{p}\vec{q}} \Psi_R(\vec{q}) +  \Psi^\dagger_R(\vec{q}) \check{t}^\dagger_{\vec{p}\vec{q}}  \Psi_L(\vec{p})  \right],
\end{equation}
where $\Psi_{L,R}(\vec{p}) = \left(\psi_\uparrow(\vec{p}), \psi_\downarrow(\vec{p}), \psi^\dagger_\downarrow(\vec{p}), -\psi^\dagger_{\uparrow}(\vec{p}) \right)^T$ are the Nambu spinors corresponding to the left and right leads, $\psi^\dagger$ and $\psi$  are the electron creation and annihilation operators, and $(\vec{p}, \vec{q})$ denote electron (hole) momenta.  We assume that the intermediate insulating layer (I) serves as a tunneling boundary, and it is therefore described by the tunneling matrix $\check{t}_{\vec{p} \vec{q}} = t_{\vec{p}\vec{q}}\check{\tau}_3$ where $\check{\tau}_3$ is the third Pauli matrix in Nambu space; the matrix element $t_{\vec{p}\vec{q}}$ obeys $\langle t_{\vec{p} \vec{q}} t^\star_{\vec{p}'\vec{q}'} \rangle =  \vert t \vert^2 \delta(\vec{p} - \vec{p}')\delta(\vec{q} - \vec{q}')$, where  averaging is carried out over the disorder distribution of tunneling amplitudes.

Assuming a phase difference $\phi$ between the superconducting leads and a normal-state boundary resistance $R_B \propto 1/\vert t \vert^2$, we have obtained the free energy of the JJ in the lowest non-trivial order in $\hat{H}_T$, and it takes the  form: 
\begin{subequations}\label{eq:F_F0_FF}
\begin{align}
F_{JJ} &= - \frac{\pi T}{8e^2 R_B}  
\sum_{\omega_n} \operatorname{Tr}\!\left[\check{g}_R(\omega_n)\check{g}_L(\omega_n) - 4 \right]=  \label{eq:F_F0_FF_a} \\[6pt]
&= \frac{I_c}{2e}\left(1 - \cos\phi\right) + F_{JJ0}. \label{eq:F_F0_FF_b}
\end{align}
\end{subequations}
Here, $\check{g}_{L}$ and $\check{g}_R$ are the quasiclassical Green's functions of the left and right leads, $T$ is the system temperature, and $\omega_n = \pi T(2n + 1)$, $n = 0,\pm1,\pm 2...$ are the fermionic Matsubara frequencies. 
The formula (\ref{eq:F_F0_FF}) does not depend on the nature of $\check{g}_{R}$ or $\check{g}_L$, and its detailed derivation is given in Supporting Information~\ref{sec:Tunneling_energy}. The first term in Eq.~(\ref{eq:F_F0_FF_b}) reproduces the Josephson coupling $E_{JJ}$ in Eq.~(\ref{eq:Usual_Josephson_energy}).  
When the BCS Green's functions with equal order parameters, $\Delta_L = \Delta_R = \Delta$, are substituted  into Eq.~(\ref{eq:F_F0_FF}), then  $F_{JJ0}=0$, so $F_{JJ} = E_{JJ}$. 

However, when the Green's functions are not equal to each other, $F_{JJ0}$ is not necessarily zero. To be more specific, we consider the system shown in Fig.~\ref{fig:setup} where $h_L = h_R = h=ba/d_S$.  Then the critical current $I_c$ and $F_{JJ0}$ can be written as:
\begin{align}
     & I_c = -\frac{\pi T}{e R_B}\sum_{\omega_n}\left[ f^2_{s} + f^2_t \cos \theta\right]  ; \label{eq:Ic}\\
     &F_{JJ0} = - \frac{\pi T}{2e^2 R_B} \sum_{\omega_n} \left[ g^2_{s} + g^2_t \cos\theta  \right. \nonumber \\
     & \left.  \qquad\qquad -\left( f^2_{s} + f^2_t\cos\theta \right) - 1\right]\; . \label{eq:F_JJ0}
\end{align}
Here, $\theta$ is the angle between $\vec{m}_R$ and $\vec{m}_{L}$;  $f_{s,t}(\omega_n) = \frac{1}{2}\left[f_0 (\omega_{n+}) \pm f_0(\omega_{n-})\right]$, $g_{s,t}(\omega_n) = \frac{1}{2}\left[g_0 (\omega_{n+}) \pm g_0(\omega_{n-})\right]$, $f_0(\omega_n) = \frac{i\Delta}{\sqrt{\omega^2_n + \Delta^2}}$, $g_0(\omega_n) = \frac{ \omega_n }{\sqrt{\omega^2_n + \Delta^2}}$, and $\omega_{n\pm} = \omega_n \pm ih$.
Using Eq.~\eqref{eq:F_JJ0}, it can be seen that $F_{JJ0} \neq 0$ for non-vanishing $\theta$. 
In the derivation of the above expressions and the results below, we have neglected proximity effects between the two superconductors of the junction, which is reasonable  in the tunneling limit. Indeed, accounting for  proximity effects does not qualitatively alter our conclusions.

\textbf{Ground state of the junction.}
To understand the consequences of  $F_{JJ0}$, let us consider the regime in which the exchange fields are weak, i.e., $h_L = h_R = h \ll (\Delta, T)$. In this case, Eq.~(\ref{eq:F_JJ0}) yields:
\begin{equation} \label{eq:F_JJ0_S}
    F_{JJ0} =  - \frac{\pi }{e^2 R_B } \frac{h^2 \left(T \sinh \left(\frac{\Delta }{T}\right)-\Delta \right) }{8 \Delta  T \cosh^2 \frac{\Delta}{2 T}} \sin^2 \frac{\theta}{2}.
\end{equation}
Since $T \sinh \frac{\Delta}{T} - \Delta > 0$, $F_{JJ0}$ is negative with a minimum at  $\theta = \pi$. In other words, it favors the antiparallel (AP) configuration of the FIs. Physically, $F_{JJ0}$ is the generalization to the case of a JJ of the classic result by de Gennes~\cite{DEGENNES1966}  for the interaction between two FIs mediated by a thin superconducting film, which has been experimentally investigated in Refs.~\cite{Moodera13,zhu2017superconducting,matsuki2025realisation}.  On the other hand, it can be shown that the critical current exhibits a minimum in the parallel (P) configuration of the FI layers (see Supporting Information~\ref{sec:Critical_Current_h}) and therefore the Josephson energy from Eq.~\eqref{eq:Usual_Josephson_energy} is minimized in the P configuration. Thus, we conclude that the AFM coupling resulting from Eq.~\eqref{eq:F_JJ0_S} competes with the  Josephson energy. This competition induces a phase transition between the AP and P configurations driven by either the temperature or the phase difference between the  superconducting leads, as shown in Figs.~\ref{fig:setup}(b) and (c).

We can further simplify the expression of the total energy to better understand what determines the JJ ground state. The latter is obtained by minimization of Eq.~\eqref{eq:F_F0_FF_b} with respect  to $\theta$ at a  fixed $\phi$.
From Eqs.~(\ref{eq:F_F0_FF_b}), (\ref{eq:Ic}), and (\ref{eq:F_JJ0}) the contributions to the free energy depending on $\theta$ can be written as follows:
\begin{equation}
F_{JJ}=a\left(1+r\cos\phi\right)\cos \theta+\ldots
\label{eq:FJJ_gen}
\end{equation}  
where  $a=-\frac{\pi T}{2e^{2}R_{B}}\sum g_{t}^{2}$, $r=-{\sum f_{t}^{2}}/{\sum g_{t}^{2}}$, with  terms independent of $\theta$ hidden in the ellipsis. Eq.~\eqref{eq:FJJ_gen} shows that $F_{JJ}(\theta,\phi)$ has extrema only at $\theta = 0,\pi$, corresponding to the P and AP configurations. Since $f_t^2>0$ and $g_t^2<0$, $a$ and $r$ are both positive parameters.  Recalling that $\cos \theta = \vec{m}_L \cdot \vec{m}_R$, the term proportional to $r$ in Eq.~\eqref{eq:FJJ_gen} resembles the phenomenological free energy of Refs.~\cite{WaintalBrouwer02,Linder11} responsible for the AP-P transition in those works.  
However, Eq.~\eqref{eq:FJJ_gen} implies that such a transition occurs only if $r > 1$ and $\pi/2 < \phi < 3\pi/2$. In our system, at low temperatures (i.e. $T \ll \Delta$), we find $r \approx 1/3$, thus the ground-state is always AP, regardless of the value of the superconducting phase difference, $\phi$.  
This result contrasts with the behavior of all-metallic spin valves studied in previous works \cite{WaintalBrouwer02,Shomali_2011} in which the AP-P transition was also obtained  at $T=0$.

\begin{figure}[tbp]
    \centering    
\includegraphics[width=0.95\linewidth]{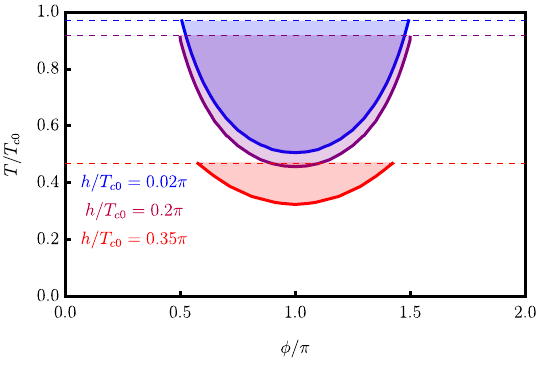}
    \caption{Zero-field phase diagrams in the $(\phi, T)$ plane illustrating the AP-P phase transition for different exchange fields $h$. The horizontal dashed lines indicate $T_c(h)$, the temperature above which superconductivity is suppressed by the exchange field. The shaded areas correspond to the parallel phase, while the unshaded regions below the dashed lines represent antiparallel alignment of the magnetic moments.}
    \label{fig:AFM_FM_phases}
\end{figure}

 It can be shown that $r$ increases monotonically with temperature.
Near $T_c$, $r \gg 1$, and therefore in Eq.~\eqref{eq:FJJ_gen} the term proportional to $r$ dominates.  In addition, for $h \ll T_c$, the free energy takes the  form:
\begin{equation} \label{eq:F_JJ_near_Tc}
\begin{aligned}
     & F_{JJ}= \frac{\pi}{e^2 R_B} \frac{h^2 \Delta ^2 }{96 T^3_{c}}   \cos \theta  \cos \phi +\ldots
\end{aligned} 
\end{equation}
Thus, the P configuration is the ground state of the junction when $\cos \phi < 0$, i.e., for $\pi/2 < \phi < 3\pi/2$.

For arbitrary values of temperature and exchange field, the phase diagrams obtained numerically are shown in Fig.~\ref{fig:AFM_FM_phases}.  
These results confirm that, while the AP phase is the ground state at low temperatures, a transition to the P phase takes place at sufficiently high temperatures for phase differences $\pi/2 < \phi < 3\pi/2$. By increasing the exchange field, the temperature at which the AP–P transition occurs is reduced.




The AP-P  phase transition can be controlled not only by temperature and phase bias, but also by an external magnetic field, $\vec{B}$. In this case, we need to include the Zeeman energy of the FI layers, which can be written as
\begin{equation} \label{eq:Zeeman}
    F_Z = - M_0\vec{B}\cdot(\vec{m}_L + \vec{m}_R),
\end{equation}
where $M_0$ is the magnetic moment of the FI films. 
For simplicity, we assume that $\vec{B}$ is aligned along the $z$-axis, i.e., $\vec{B}\propto \vec{z}$. We disregard the Meissner effect since in thin, dirty films the London penetration depth $\lambda_L$ is the largest length scale, and therefore $d_S \ll \xi \ll \lambda_L$, meaning that screening corrections $O(d_S/\lambda_L)$ are negligible. In addition, we also neglect the paramagnetic effect of the external magnetic field on the superconducting leads, since $\mu_B B \ll (\Delta,\ h)$, where $\mu_B$ is the Bohr magneton.

\begin{figure}[htbp]
    \centering    \includegraphics[width=0.95\linewidth]{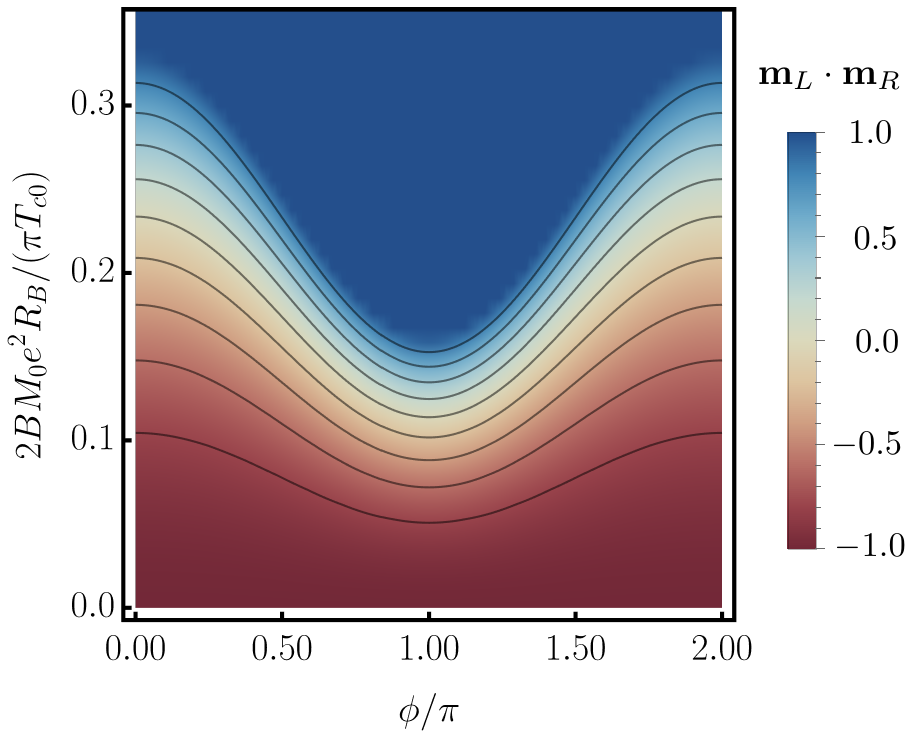}
    \caption{Phase diagram for  our device with $h/T_{c0}=0.2\pi$ at absolute temperature $T=0.05 T_{c0}$ as a function of the superconducting phase $\phi$ and the external magnetic field $B$ applied along $\vec{z}$ axis.  
    }
    \label{fig:phaseB}
\end{figure}

Fig.~\ref{fig:phaseB} shows the effect of an external magnetic field on the system at $T = 0.05 T_{c0}$. With increasing field strength, the system undergoes a continuous AP-P transition, as evidenced by the gradual growth of the net magnetic moment, $M_0 \vert \vec{m}_L + \vec{m}_R \vert$. This behavior is natural, since the external field tends to align the magnetic moments of the two FI layers, thereby allowing the Josephson energy to overcome the competing $F_{JJ0}$ contribution. As a result, the phase diagram again exhibits a P region centered around $\phi \approx \pi$, even at low temperatures. 

\textbf{Switching dynamics of the JJ:} The presence of two stable configurations of the FI magnetizations, determined either by the superconducting phase difference $\phi$ or by the external magnetic field $\vec{B}$, suggests a promising route toward the construction of a memory device, where the two distinct states can be assigned as logical "0" and "1". In our convention, the AP alignment of the FI layers corresponds to the logical "0", while the P alignment is assigned to the logical "1". Below, we investigate the possibility of switching between these two states in our JJ by applying time-dependent phase or magnetic biases.

For this purpose, we employ the Landau–Lifshitz–Gilbert (LLG) equation~\cite{landau2013course}, with the system’s free energy $F = F_{JJ} + F_Z + F_M$, consisting of three contributions: The tunneling energy $F_{JJ}$ from Eq.~(\ref{eq:F_F0_FF}), the Zeeman energy $F_Z$ from Eq.~(\ref{eq:Zeeman}), and the easy-plane magnetic anisotropy energy, typical of EuS FI films~\cite{Story99},
\begin{equation} \label{eq:F_Anisotropy}
    F_M = \frac{K }{2}\left(m_{Lx}^2 + m_{Rx}^2 \right).
\end{equation}
Here $K>0$ is the magnetic anisotropy constant, and $m_{(L,R)x}$ are the $x$-components of the FI  magnetization-direction unit vectors, $\vec{m}_L$ and $\vec{m}_R$.  
\begin{figure}[htbp]
    \centering
    \includegraphics[width=0.95\linewidth]{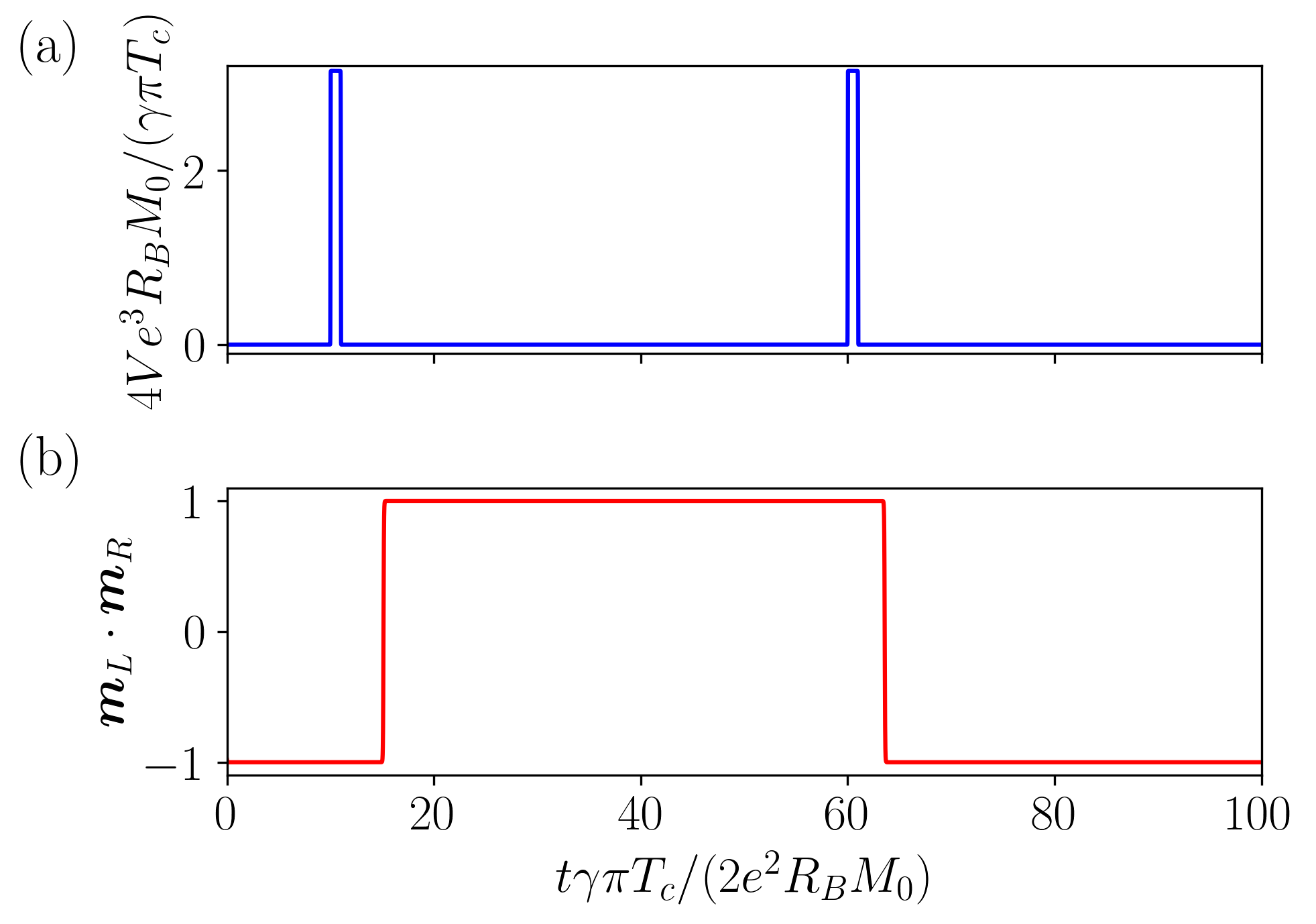}
    \includegraphics[width=0.95\linewidth]{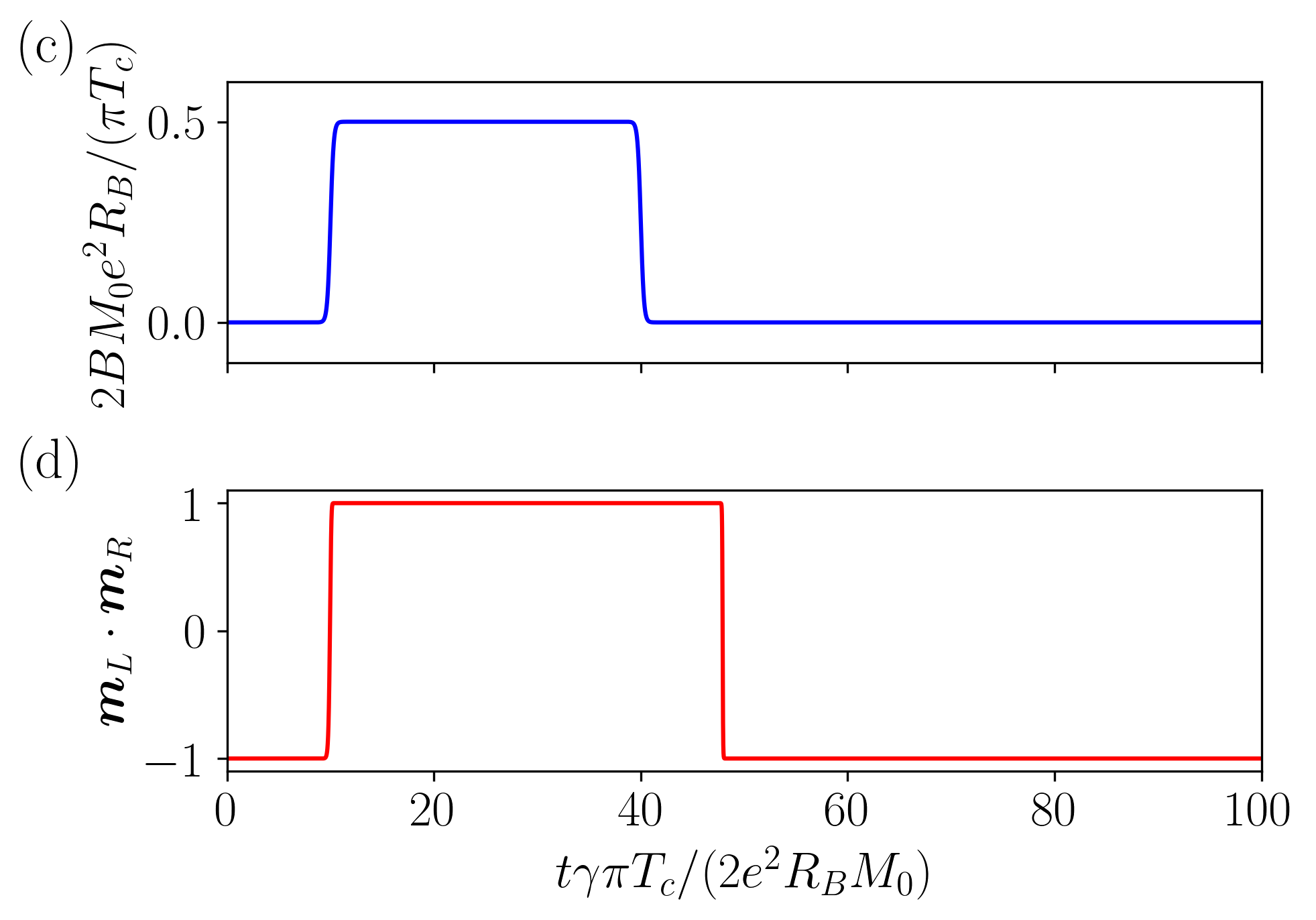}
    \caption{Switching between AP and P configurations induced by half-flux-quantum (HFQ) voltage pulses (a,b) and by a magnetic field pulse (c,d). The system parameters correspond to the EuS/Al/AlO$_x$/Al/EuS junction discussed in the main text. Two voltage pulses are applied at $\tfrac{t_0 \gamma \pi T_c}{2e^2 R_B M_0} = 10$ and $\tfrac{t_1 \gamma \pi T_c}{2e^2 R_B M_0} = 60$, each with duration $\delta t = \tfrac{2e^2 R_B M_0}{\gamma \pi T_c} \approx 5.5\ \mu\text{s}$ and amplitude $V_p = \tfrac{\gamma \pi T_c}{4 e^3 R_B M_0} \times \tfrac{\pi}{\delta t} \approx 34\ \mu\text{V}$. The phase difference is computed from the second Josephson relation, $\dot{\phi}(t)/(2e) = V(t)$, with $\phi(-\infty) = 0$, yielding $\phi(t) = \int_{-\infty}^t 2eV(t)\,dt$. The magnetic field pulse is applied along the $\vec{y}$ axis at $\tfrac{t_0 \gamma \pi T_c}{2e^2 R_B M_0} = 10$, with duration $\delta t = 30 \times \tfrac{2e^2 R_B M_0}{\gamma \pi T_c}$ and amplitude $B_p = 0.5 \times \tfrac{\pi T_c}{2e^2 R_B M_0} \approx 0.54\ \mu\text{T}$.}
    \label{fig:Switching}
\end{figure}
The LLG equations read
\begin{equation} \label{eq:LLG_L_R}
    \frac{d \vec{m}_{L,R}}{d t} 
    = \gamma \, \vec{H}_{L,R} \times \vec{m}_{L,R} 
    + \alpha \, \vec{m}_{L,R} \times \frac{d \vec{m}_{L,R}}{d t},
\end{equation}
where $\gamma$ is the electron gyromagnetic ratio, $\vec{H}_{L,R}$ are the effective magnetic fields acting on the left and right FI layers, and $\alpha$ is the Gilbert damping coefficient. The effective fields are obtained from:
\begin{equation}
    \vec{H}_{L,R} = - \frac{1}{M_0}\frac{\partial F}{\partial \vec{m}_{L,R}}, 
    \qquad \vec{m}_L \cdot \vec{m}_R = \cos \theta,
\end{equation}
which yields 
\begin{equation} \label{eq:H_LR}
    \vec{H}_{L,R} = - \frac{\pi}{e^2 R_B} \frac{h^2 \Delta^2 \cos \phi}{96 T^3_{c} M_0} \vec{m}_{R,L}  + \vec{B} - \frac{K}{M_0}  m_{L,Rx} \vec{x}.
\end{equation}
For the sake of simplicity, we restrict our analysis to temperatures close to the critical temperature, $T_c - T \ll T_c$. In this regime, the tunneling energy $F_{JJ}$ can be approximated by Eq.~\eqref{eq:F_JJ_near_Tc}, which leads  to the first term in Eq.~\eqref{eq:H_LR}. 


To make the discussion more realistic, we focus on an EuS/Al/AlO$_x$/Al/EuS heterostructure, a material combination widely used in experiments \cite{moodera1988electron,moodera1988electron,moodera1988electron,Strambini17,Giazotto18,Hijano21}. We assume a junction area in the $\vec{y}$–$\vec{z}$ plane (see Fig.~\ref{fig:setup}(a)) of $\mathcal{A} = 100\ \mu\text{m} \times 100\ \mu\text{m}$.
For the barrier resistance per unit area, we choose $R_A = 30\ \Omega \cdot \mu\text{m}^2$ \cite{Lotkhov06}, corresponding to $R_B = R_A/\mathcal{A} = 3.0 \times 10^{-3}\ \Omega$.  
For the superconducting layers, we take $d_S = 10\ \text{nm}$, $\Delta = 0.24\ \text{meV}$, $T_c = 1.3\ \text{K}$, and $h = 0.12\ \text{meV}$ \cite{Hijano21}.  
The magnetic anisotropy constant of the EuS layers is set to $K/(\mathcal{A} d_F) = 0.7\times 10^6\ \text{J}/\text{m}^3$ in accordance with Ref.~\cite{Story99}, where $d_F$ denotes the FI thickness along the $\vec{x}$ direction.  
The magnetic moment density is estimated as $M_0/(\mathcal{A} d_F) = 1.2\times10^6\ \text{A}/\text{m}$, using the facts that EuS crystallizes in the rocksalt structure with a lattice constant $a = 5.97\times 10^{-10}\ \text{m}$ and that each Eu$^{2+}$ ion carries spin $S = 7/2$ \cite{Story99}. The Gilbert damping is taken as $\alpha = 10^{-2}$, consistent with the order of magnitude reported in Ref.~\cite{aguilar2023magnon}.


The switching between the AP and P configurations in two scenarios --- by applying voltage pulses or by applying a time-dependent magnetic field --- is shown in Fig.~\ref{fig:Switching}. We employ rectangular voltage pulses satisfying $\int dt\, 2eV(t) = \pi$. Unlike conventional SFQ pulses, which produce $2\pi$ phase jumps~\cite{Soloviev2017}, these pulses induce only $\pi$ shifts in $\phi$ and are therefore referred to as half-flux-quantum (HFQ) pulses. Such pulses can be realized, for example, by using JJs with a CPR dominated by the second harmonic, $I_S \propto \sin 2\phi$~\cite{Soloviev2021AllJJ,Mitrovic25}. For magnetic-field-induced switching, we apply rectangular field pulses along the $\vec{y}$ axis. As illustrated in Figs.~\ref{fig:Switching}(a)--(d), the characteristic time scale for magnetization dynamics in response to external stimuli is $\frac{2 e^2 R_B M_0}{\gamma \pi T_c} \approx 5.5\ \mu\text{s}$. This represents an improvement of three to four orders of magnitude over the characteristic writing time per bit ($\sim 10\ \text{ms}$) for random single-bit operations in magnetic hard drives. 

The AP and P configurations are characterized by different critical currents, $I^{AP}_{c}$ and $I^P_{c}$, which satisfy the inequality $I^P_{c} < I^{AP}_{c}$. This distinction enables non-destructive electrical readout of the EuS/Al/AlO$_x$/Al/EuS memory element using current pulses. As noted above, the lifetime of the FI layers in the AP or P configuration is about $1\ \mu\text{s}$, which is extremely long compared to the typical superconducting time scale of the order of $1/\Delta \sim 10^{-12}\ \text{s}$. Within this long-lived interval, a current $I = (I^{P}_{c} + I^{AP}_{c})/2$ can be applied to the junction. If such a pulse generates a voltage response, the junction resides in the AP state; if not, it is in the P state. Owing to the large separation of timescales between magnetization dynamics and phase evolution, properly engineered readout pulses can probe the junction without disturbing its state.

As shown in the Supporting Information~\ref{sec:Switching_Stability_Analisys}, switching is always possible for sufficiently large easy-plane anisotropy $K$ and finite Gilbert damping. Indeed, for temperatures and exchange fields where the AP–P transition is possible, once the stable regime is flipped, the system relaxes to its new equilibrium state.

\textbf{Conclusions.} In this work, by analyzing  the tunneling between in a Josephson junction of two spin-split superconductors, we have identified  two distinct contributions to Josephson energy. The first contribution takes the form of the conventional Josephson term, $E_{JJ}(\phi)$, and corresponds to the work performed by a current source to establish a supercurrent $I_c \sin\phi$ in a junction initially at zero current. The second term, $F_{JJ0}$, captures the asymmetry between the superconducting leads. In junctions formed by ferromagnetic insulator (FI) / superconductor  hetero-structures, resulting in two spin-split superconductors separated by an insulating layer, this asymmetry gives rise to an effective antiferromagnetic coupling between the magnetizations of the FI layers. The latter is mediated by the superconducting condensate tunneling through the insulating barrier of the junction. This coupling competes with the previously mentioned Josephson term, which favors parallel alignment of the magnetizations of the FIs.
The competition between the two contributions allows one to switch the FI magnetizations between parallel and antiparallel configurations.  The switching takes place in FIs with large magnetic easy-plane anisotropy energies and occurs at finite temperatures and  for phase-differences satisfying  $\pi/2<\phi<3\pi/2$. We believe that our results open a novel pathway towards all-electrical, hybrid superconducting-magnetic memory elements based on, e.g. EuS/Al/AlO$_x$/Al/EuS heterostructures, which would operate without relying on spin–orbit coupling or external magnetic fields.

\textit{Acknowledgments.} A. M. and F. S. B. acknowledge fruitful discussions with Norman Birge and Max Ilyn, and financial support from the European Union’s Horizon Europe research and innovation program under grant agreement No. 101130224 (JOSEPHINE). F.S.B  acknowledges financial support from the Spanish MCIN/AEI/10.13039/501100011033 
through the grants PID2023-148225NB-C31 and TED2021-130292B-C42. C.-H.H has been supported by EU's HORIZON-RIA Programme under Grant No. 101135240 (JOGATE).
M.A.C. has been supported by the Spanish MCIN/AEI/10.13039/501100011033 through Grant 
No. PID2023-148225NB-C32 (SUNRISE).

\bibliographystyle{apsrev4-2}
\bibliography{refs}
\clearpage
\onecolumngrid
\appendix

\section{Tunneling energy derivation}
\label{sec:Tunneling_energy}

The influence of tunneling coupling given by Eq.~(\ref{eq:tunneling_coupling}) on the thermodynamic properties of superconductors can be calculated with the help of the Matsubara technique. The partition function of the Josephson junction reads as follows,
\begin{equation} \label{eq:Z}
    Z = Z_0 \left\langle \operatorname{T}_\tau \exp \left\{ - \int_{0}^{\beta} d\tau'\ \hat{H}_T(\tau') \right\} \right\rangle_0, 
\end{equation}
where $\tau'$ is the imaginary time, $\beta = \frac{1}{T}$, $T$ is the system temperature. $Z_0$ is the partition function of two isolated superconductors described by $\hat{H}_L$ and $\hat{H}_R$ in Eq.~(\ref{eq:H_total}), and the averaging $\langle \rangle_0$ is carried out over an ensemble of two non-interacting condensates in the left and right leads.

With the help of the relation $F = -T \log Z$ and the Wick theorem, we calculate the free energy difference between two coupled superconductors and two isolated superconductors:
\begin{equation} \label{eq:F_F0}
\begin{aligned}
    &\delta F= F - F_0 =  \int_{0}^{\beta}d\tau' \int_{0}^{\beta}d\tau'' \sum_{\vec{p}'\vec{p}''\vec{q}'\vec{q}''} t_{\vec{p}'\vec{q}'} t^\star_{\vec{p}''\vec{q}''} \times  \operatorname{Tr}\left\{ \check{G}_R(\tau',\vec{q}';\ \tau'',\vec{q}'')  \check{G}_L(\tau'',\vec{p}'';\ \tau',\vec{p}')  \right\}.
\end{aligned}
\end{equation}
Here, $\check{G}_{L,R}(\tau,\vec{q};\ \tau',\vec{q}') = -\check{\tau}_3 \langle \Psi(\tau,\vec{q}) \Psi^\dagger(\tau',\vec{q}') \rangle_{0}$ are the Gor'kov  functions of the S leads. Using the tunneling matrix property, $\langle t_{\vec{p} \vec{q}} t^\star_{\vec{p}'\vec{q}'} \rangle =  \vert t \vert^2 \delta(\vec{p} - \vec{p}')\delta(\vec{q} - \vec{q}')$, we perform the averaging over the disorder in the tunneling amplitudes in Eq.~(\ref{eq:F_F0}) and find following  relation for the tunneling energy:
\begin{equation} \label{eq:F_F0_F}
    \delta F = - \frac{\pi T}{8e^2 R_B}  \sum_{\omega_n} \operatorname{Tr}\left\{ \check{g}_R(\omega_n) \check{g}_L(\omega_n)\right\},
\end{equation}
where $\omega_n = \pm \pi T(2n +1)$ are the fermionic Matsubara frequencies ($n = 0, \pm 1,...$),  $\check{g}_{L,R} = \frac{i}{\pi} \int d\xi\ \check{G}_{L,R}$ are the quasiclassical Green's functions of the leads, $\xi = \frac{p^2}{2m} - E_F$ is the quasiparticle energy measured from the Fermi energy, $E_F$, and $R_B = \left(8 \pi e^2 \vert t \vert^2 V_L V_R N_{0L} N_{0R} \right)^{-1}$ is the normal state boundary resistance of the middle I layer, $V_{L,R}$ and $N_{0L,R}$ are the volumes and densities of states at the Fermi energy of the left and right superconductors.  

We assume the phase bias of our junction that results in the following gauge transformation of the Green's functions: $\check{g}_{L,R} \to e^{\mp i\phi \check{\tau}_3/2} \check{g}^{(0)}_{L,R}  e^{\pm i\phi \check{\tau}_3/2}$, where $\check{g}^{(0)}_{L,R}$ are the quasiclassical BCS Green's functions of the leads corresponding to $\phi = 0$. 

The energy difference $\delta F$ in Eq.~(\ref{eq:F_F0_F}) contains a divergent term that originates from the normal components of the Green's functions $\check{g}_{L,R}$. 
 To remove this divergence, we perform the renormalization of $\delta F$ by subtracting the normal-state energy difference corresponding to $\check{g}_{L,R} = \check{\tau}_3 \operatorname{sgn}{\omega_n}$, so we end up with
 \begin{align} \label{eq:F_F0_FF_Appendix}
F_{JJ} &= \delta F - \delta F(\Delta_{L,R}=0) = - \frac{\pi T}{8e^2 R_B}  
\sum_{\omega_n} \operatorname{Tr}\!\left[\check{g}_R(\omega_n)\check{g}_L(\omega_n) - 4 \right] = \frac{I_c}{2e}\left(1 - \cos\phi\right) + F_{JJ0}. 
\end{align}
This equation (\ref{eq:F_F0_FF_Appendix}) corresponds to Eqs.~\ref{eq:F_F0_FF}(a) and (b) of the main text.

\section{Critical currents for parallel and antiparallel configurations}
\label{sec:Critical_Current_h}

\begin{figure}
    \centering
    \includegraphics[width=0.32\linewidth]{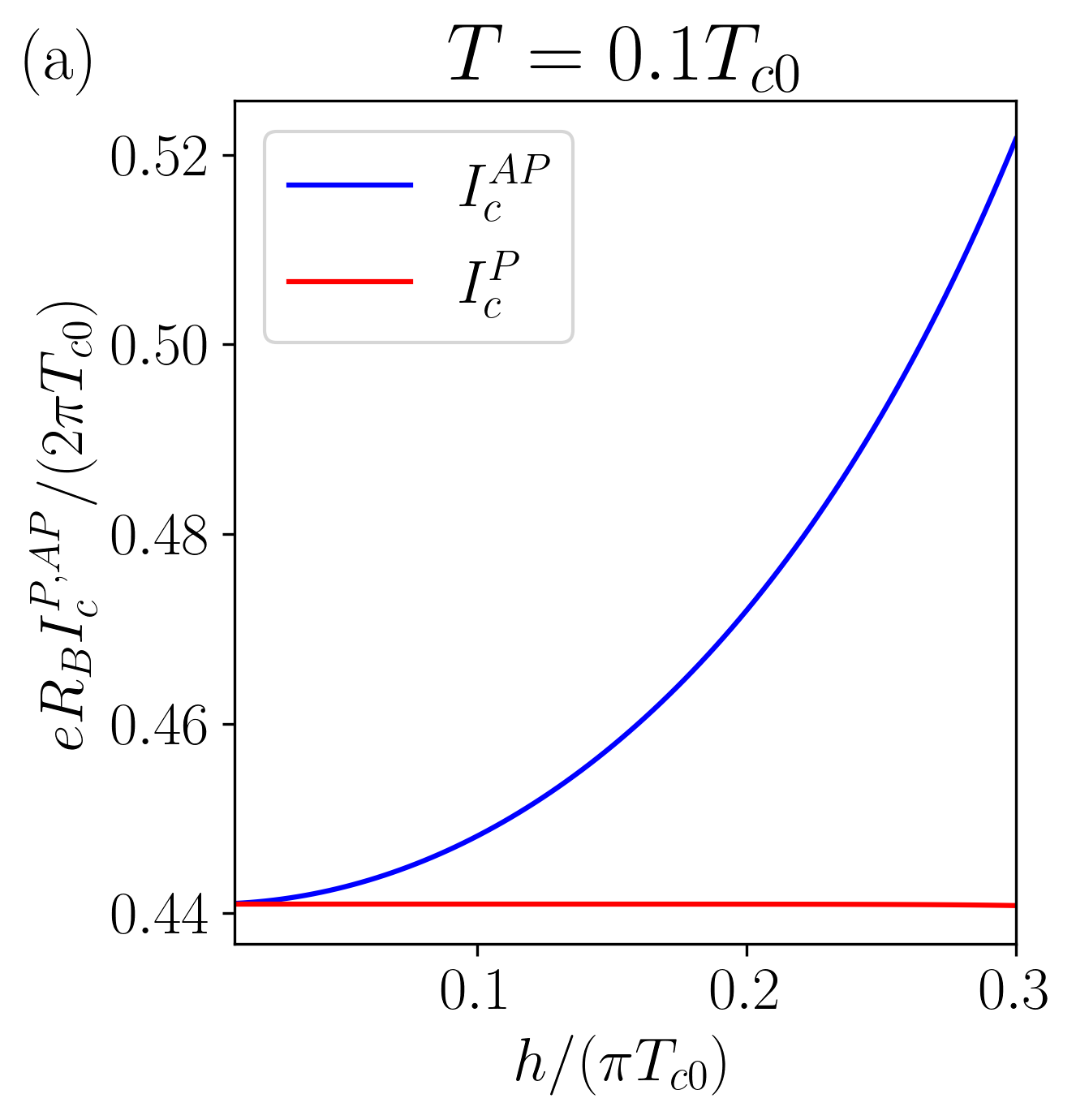}
    \includegraphics[width=0.32\linewidth]{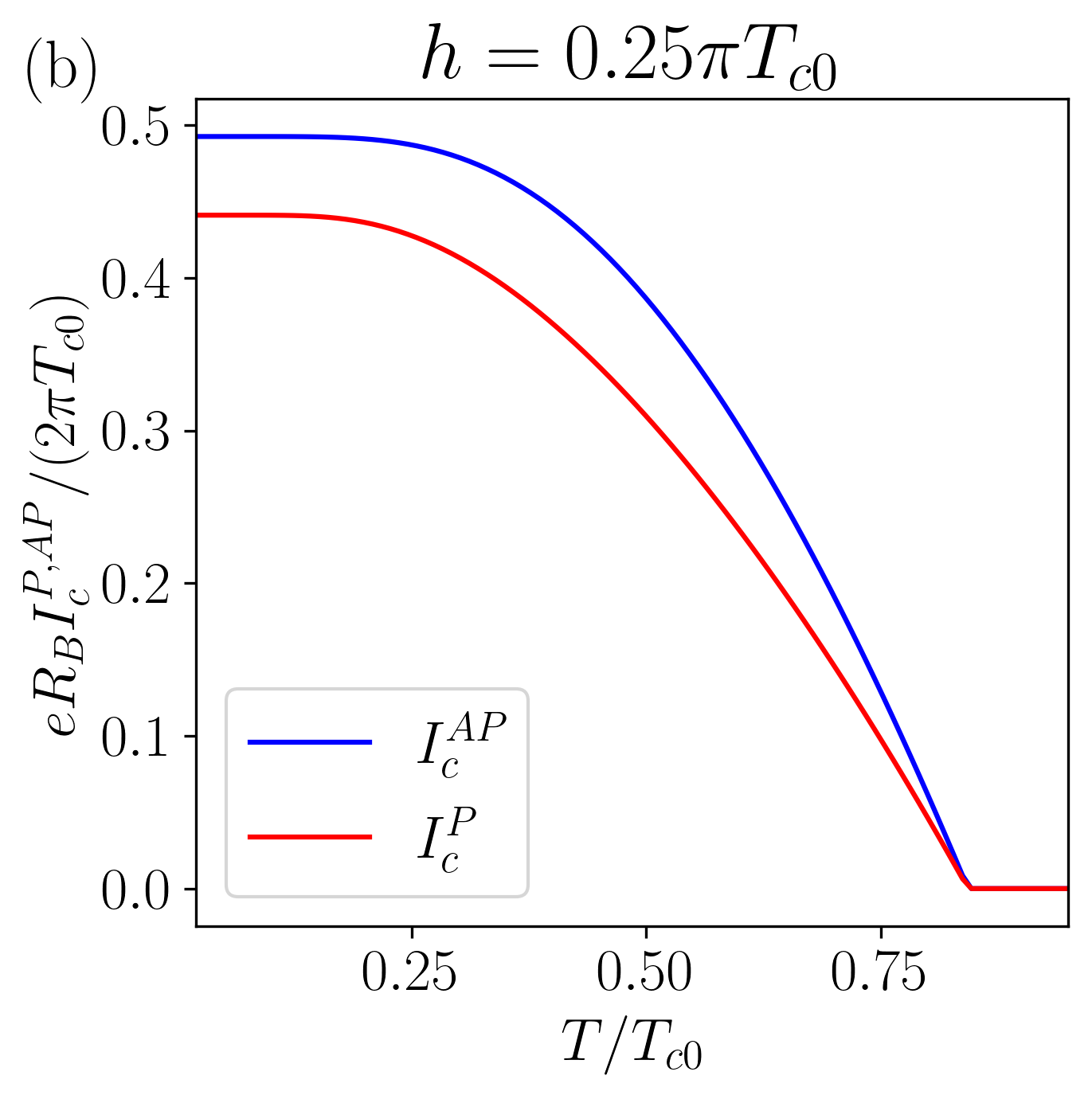}
    \caption{Critical current dependencies on the spin-splitting field $h$ and temperature $T$ for parallel (a) and antiparallel (b) configurations of the FI magnetic moments according to Eqs.~(\ref{eq:Ic_P}) and (\ref{eq:Ic_AP}). }
    \label{fig:Ic_h_t}
\end{figure}

The critical current of the junction under consideration is given by Eq.~(\ref{eq:Ic}) of the main text,
\begin{equation}
    I_c = -\frac{\pi T}{e R_B}\sum_{\omega_n}\left[ f^2_{s} + f^2_t \cos \theta\right].
\end{equation}
With the help of the singlet and triplet anomalous Green's functions of spin-split superconductors,
\begin{equation}
    f_{s,t}(\omega_n) = \frac{1}{2}\left[f_0 (\omega_{n+}) \pm f_0(\omega_{n-})\right],\qquad \omega_{n\pm} = \omega_n \pm ih,\qquad f_0(\omega_n) = \frac{i\Delta}{\sqrt{\omega^2_n + \Delta^2}},
\end{equation}
we write the critical currents in the parallel (P) and antiparallel (AP) cases as follows:
\begin{align}
    &I^{P}_c =  \frac{2\pi \Delta^2 T}{e R_B} \sum_{\omega_n > 0}  \frac{\omega^2_n - h^2 + \Delta^2}{\left(\omega^2_n - h^2 + \Delta^2\right)^2 + 4 \omega^2_n h^2}, \label{eq:Ic_P}\\
    &I^{AP}_c =  \frac{2 \pi \Delta^2 T}{e R_B} \sum_{\omega_n > 0}  \frac{1}{\omega^2_n - h^2 + \Delta^2 }. \label{eq:Ic_AP}
\end{align}
We compute the critical currents using Eqs.~(\ref{eq:Ic_P}) and (\ref{eq:Ic_AP}) and present them in Figs.~\ref{fig:Ic_h_t}(a) and (b). From these equations, one can check that  $I^{P}_c < I^{AP}_c$, as stated in the main text. 

The Matsubara sums in Eqs.~(\ref{eq:Ic_P}) and (\ref{eq:Ic_AP}) are carried out numerically, as in the rest of this paper, with the help of the trick discussed in Ref.~\cite{Virtanen2025}.    The dependence of $\Delta$ on $h$ is also taken into account; see the Supporting Information~\ref{sec:Delta_h} for details.  

\section{\texorpdfstring{$\Delta(h,T)$}{} dependence in the superconducting leads}
\label{sec:Delta_h}

The superconducting leads considered in this work are subjected to an effective homogeneous exchange field, $h = b a / d_S$. To determine the corresponding order parameter $\Delta = \Delta(h, T)$, we start from the free-energy density, given by the standard BCS expression~\cite{coleman2015introduction},
\begin{equation} \label{eq:F_S}
F_S = \frac{|\Delta|^2}{V}
      - 2T \sum_{\vec{k},\sigma}
      \ln\!\left[ 2\cosh\!\left(\frac{E_{\vec{k}}^\sigma}{2T}\right) \right],
\end{equation}
where $E_{\vec{k}}^\sigma = \sqrt{\xi_{\vec{k}}^2 + |\Delta|^2} + \sigma h$ is the quasiparticle energy for spin $\sigma$ and momentum $\vec{k}$, measured relative to the Fermi energy $E_F$. Here, $\xi_{\vec{k}} = \frac{|\vec{k}|^2}{2m} - E_F$ is the single-particle dispersion, $m$ is the electron mass, and $V$ denotes the BCS coupling constant.

The self-consistency equation for the order parameter follows from the stationarity condition $\partial F_S / \partial \Delta^\ast = 0$, which yields
\begin{equation} \label{eq:Self-Consistency}
\frac{2}{V} = \sum_{\vec{k},\sigma}
  \tanh\!\left(\frac{E_{\vec{k}}^\sigma}{2T}\right)
  \frac{1}{E_{\vec{k}}^\sigma|_{h=0}}.
\end{equation}
At zero temperature, Eq.~\eqref{eq:Self-Consistency} reduces to
\begin{equation} \label{eq:Delta_00}
\frac{2}{V}
= \sum_{\vec{k}} \frac{2}{\sqrt{\xi_{\vec{k}}^2 + \Delta_0^2}},
\end{equation}
where $\Delta_0 = \Delta(h=0, T=0)$ is the zero-temperature superconducting gap in the absence of exchange splitting.

Substituting Eq.~\eqref{eq:Delta_00} into Eq.~\eqref{eq:F_S}, we eliminate the explicit $V$-dependence and obtain
\begin{align} \label{eq:F_S_final}
F_S = \sum_{\vec{k}}\!\left[
    \frac{|\Delta|^2}{\sqrt{\xi_{\vec{k}}^2 + \Delta_0^2}}
    - 2T \sum_{\sigma}
    \ln\!\left(2\cosh\!\frac{E_{\vec{k}}^\sigma}{2T}\right)
\right].
\end{align}
Finally, the order parameter $\Delta(h, T)$ is obtained by minimizing $F_S$ with respect to $\Delta$ at fixed $h$ and finite $T$.

\section{Stability analysis of P and AP equilibrium configurations}
\label{sec:Switching_Stability_Analisys}

To analyze the magnetization dynamics, it is natural to define a characteristic time scale. Since this dynamics is driven by the Josephson effect, whose typical energy scale is $F_{JJ} \propto \tfrac{\pi T_c}{2e^2 R_B}$, the corresponding magnetic time scale may be determined as the inverse of a characteristic frequency corresponding to $\gamma \left \vert \vec{H}_{L,R} \right\vert \sim \gamma \left \vert \frac{1}{M_0} \frac{\partial F_{JJ}}{\partial \vec{m}_{L,R}} \right\vert \sim \frac{\gamma F_{JJ}}{M_0}$, that is, $(\gamma F_{JJ}/M_0)^{-1}$. Accordingly, we introduce the dimensionless time variable as $\tilde{t} = \tfrac{t \gamma \pi T_c}{2e^2 R_B M_0}$. Having this in mind, we start from the effective magnetic field given by Eq.~(\ref{eq:H_LR}) of the main text and rewrite it in  a form which helps to introduce our time units, 
\begin{equation} \label{eq:H_LR_appendix}
\begin{aligned}
    &\vec{H}_{L,R} = - \frac{\pi}{e^2 R_B} \frac{h^2 \Delta^2 \cos \phi}{96 T^3_{c} M_0} \vec{m}_{R,L}  + \vec{B} - \frac{K}{M_0}  m_{L,Rx} \vec{x} = \\ 
    &\qquad \qquad \qquad\qquad=\frac{\pi T_c}{2e^2 R_B M_0}\left[  -\frac{h^2 \Delta^2 }{48 T^4_{c} }  \vec{m}_{R,L}  \cos \phi + \frac{2 \vec{B} M_0 e^2 R_B} {\pi T_c} -  \frac{2e^2 R_B K}{\pi T_c} m_{L,Rx} \vec{x}  \right].
\end{aligned}
\end{equation}
From Eq.~(\ref{eq:H_LR_appendix}), we see that the natural dimensionless forms for the phase-induced torque amplitude, $G$, magnetic field, $\tilde{\vec{B}}$, and the anisotropy $\tilde{K}$ are as follows,
\begin{equation}
    G = \frac{h^2 \Delta^2}{48 T^4_c},\qquad \tilde{\vec{B}} = \frac{2 \vec{B} M_0 e^2 R_B} {\pi T_c},\qquad \tilde{K} = \frac{2e^2 R_B K}{\pi T_c}.
\end{equation}

With the help of our dimensionless units $\tilde{t}$, $G$, $\tilde{\vec{B}}$, $\tilde{K}$, the LLG equations for the left and right magnetic moments read as 
\begin{equation} \label{eq:LLG_LR}
\begin{aligned}
    \frac{d \vec{m}_{L,R}}{d \tilde{t} } 
    = \left[- G \vec{m}_{R,L} \cos \phi + \tilde{\vec{B}} - \tilde{K} m_{L,Rx}\vec{x}\right] \times \vec{m}_{L,R} 
    + \alpha \, \vec{m}_{L,R} \times \frac{d \vec{m}_{L,R}}{d \tilde{t}}.
\end{aligned}
\end{equation}

The phase difference $\phi$ evolves according to the second Josephson relation, $\tfrac{1}{2e}\tfrac{d\phi}{dt} = V(t)$, and is thus governed by the amplitude and duration of voltage pulses $V(t)$. Since these pulses typically last only a few picoseconds \cite{Soloviev2017,Soloviev2021AllJJ,Mitrovic25} -- much shorter than any characteristic magnetization dynamics -- we neglect the detailed time profile of $\phi$ and instead determine the equilibrium orientations in the LLG equations [Eq.~(\ref{eq:LLG_LR})] for a fixed $\phi$. For that, we employ the angular representation for $\vec{m}_L$ and $\vec{m}_R$: 
\begin{equation} \label{eq:Angles}
    \vec{m}_L = \begin{pmatrix}
        \cos \theta_L\\
        \sin\theta_L \cos\varphi_L\\
        \sin\theta_L \sin\varphi_L
    \end{pmatrix}, \qquad \vec{m}_R = \begin{pmatrix}
        \cos \theta_R\\
        \sin\theta_R \cos\varphi_R\\
        \sin\theta_R \sin\varphi_R
    \end{pmatrix}.
\end{equation}
We insert Eq.~(\ref{eq:Angles}) into Eq.~(\ref{eq:LLG_LR}) and obtain the following system of equations:
\begin{align}
    &(1 + \alpha^2)\frac{d \theta_L}{d \tilde{t}} = - \tilde{H}_{yL} \sin \varphi_L + \tilde{H}_{zL} \cos\varphi_L + \alpha \left[\tilde{H}_{yL}\cos\theta_L \cos \varphi_L + \tilde{H}_{zL}\cos\theta_L \sin\varphi_L - \tilde{H}_{xL}\sin\theta_L\right], \label{eq:LLG_Angles_1}\\
   & (1 + \alpha^2) \sin\theta_L \frac{d \varphi_L}{d \tilde{t}} = \tilde{H}_{xL} \sin\theta_L  - \tilde{H}_{yL} \cos\theta_L  \cos \varphi_L  - \tilde{H}_{zL} \cos\theta_L \sin \varphi_L + \alpha\left[\tilde{H}_{zL} \cos\varphi_L - \tilde{H}_{yL}\sin\varphi_L\right];\\
   &(1 + \alpha^2)\frac{d \theta_R}{d \tilde{t}} = - \tilde{H}_{yR} \sin \varphi_R + \tilde{H}_{zR} \cos\varphi_R  + \alpha \left[\tilde{H}_{yR}\cos\theta_R \cos \varphi_R + \tilde{H}_{zR}\cos\theta_R \sin\varphi_R - \tilde{H}_{xR}\sin\theta_R\right],\\
   & (1 + \alpha^2) \sin\theta_R \frac{d \varphi_R}{d \tilde{t}} = \tilde{H}_{xR} \sin\theta_R  - \tilde{H}_{yR} \cos\theta_R  \cos \varphi_R  - \tilde{H}_{zR} \cos\theta_R \sin \varphi_R + \alpha\left[\tilde{H}_{zR} \cos\varphi_R - \tilde{H}_{yR}\sin\varphi_R\right]. \label{eq:LLG_Angles_4}
\end{align}
Here, $\tilde{\vec{H}}_{L,R} = - G \vec{m}_{R,L}\cos\phi + \tilde{\vec{B}} - \tilde{K}m_{L,Rx}\vec{x}$. For the sake of simplicity, we put $\tilde{\vec{B}} = 0$ for the stability analysis  of the equilibrium configurations. After lengthy algebra, we simplify Eqs.~(\ref{eq:LLG_Angles_1})--(\ref{eq:LLG_Angles_4}) and find:
\begin{align}
     &(1 + \alpha^2)\frac{d \theta_L}{d \tilde{t}} =  G \cos \phi \sin \theta_R \sin (\varphi_L - \varphi_R)  + \nonumber \\
     & \qquad \qquad\qquad\qquad + \alpha \left[G \cos\phi \left[\cos\theta_R \sin\theta_L - \sin\theta_R \cos\theta_L \cos(\varphi_R - \varphi_L)\right] + \tilde{K}\cos\theta_L\sin\theta_L \right], \label{eq:LLG_Angles_a}\\
   & (1 + \alpha^2) \sin\theta_L \frac{d \varphi_L}{d \tilde{t}} = - (\tilde{K}\cos\theta_L + G \cos\phi \cos\theta_R)\sin\theta_L + G \cos\phi \sin\theta_R \cos\theta_L \cos(\varphi_R-\varphi_L) + \nonumber \\ 
   &\qquad\qquad\qquad\qquad + \alpha G \cos\phi \sin \theta_R \sin(\varphi_L - \varphi_R);\\
   &(1 + \alpha^2)\frac{d \theta_R}{d \tilde{t}} =   G \cos \phi \sin \theta_L \sin (\varphi_R - \varphi_L)   + \nonumber \\
     & \qquad \qquad\qquad\qquad + \alpha \left[G \cos\phi \left[\cos\theta_L \sin\theta_R - \sin\theta_L \cos\theta_R \cos(\varphi_L - \varphi_R)\right] + \tilde{K}\cos\theta_R\sin\theta_R \right] ,\\
   & (1 + \alpha^2) \sin\theta_R \frac{d \varphi_R}{d \tilde{t}} =  - (\tilde{K}\cos\theta_R + G \cos\phi \cos\theta_L)\sin\theta_R + G \cos\phi \sin\theta_L \cos\theta_R \cos(\varphi_L-\varphi_R) + \nonumber \\ 
   &\qquad\qquad\qquad\qquad   + \alpha G \cos\phi \sin \theta_L \sin(\varphi_R - \varphi_L). \label{eq:LLG_Angles_d}
\end{align}
One can check  that the system of Eqs.~(\ref{eq:LLG_Angles_a})--(\ref{eq:LLG_Angles_d}) has two equilibrium directions:
\begin{equation}
\begin{pmatrix}
    \theta_L\\
    \varphi_L\\
    \theta_R\\
    \varphi_R
\end{pmatrix}^{(1)} = \begin{pmatrix}
    \frac{\pi}{2}\\
    \varphi\\
    \frac{\pi}{2}\\
    \varphi
\end{pmatrix} \qquad \text{ and }\qquad \begin{pmatrix}
    \theta_L\\
    \varphi_L\\
    \theta_R\\
    \varphi_R
\end{pmatrix}^{(2)} = \begin{pmatrix}
    \frac{\pi}{2}\\
    \varphi\\
    \frac{\pi}{2}\\
    \varphi + \pi
\end{pmatrix},
\end{equation}
which correspond to parallel [(1)] and antiparallel [(2)] configurations of the FI moments. Here, $\varphi$ is an arbitrary real number reflecting the rotation symmetry inside the easy planes. Without loosing the generality, we may put $\varphi = 0$.

We expand the system of Eqs.~(\ref{eq:LLG_Angles_a})--(\ref{eq:LLG_Angles_d}) near the parallel configurations of the FI moments up to linear order, $\theta_{L,R} = \tfrac{\pi}{2} + \delta \theta_{L,R}$, $\varphi_{L,R} =  \delta \varphi_{L,R}$:
\begin{align}
     &(1 + \alpha^2)\frac{d \delta\theta_L}{d \tilde{t}} =  G \cos \phi  (\delta\varphi_L - \delta\varphi_R) - \alpha \left[G \cos\phi \left[\delta \theta_R - \delta\theta_L \right] + \tilde{K} \delta\theta_L \right], \label{eq:LLG_Angles_aP}\\
   & (1 + \alpha^2)  \frac{d \delta \varphi_L}{d \tilde{t}} = \tilde{K} \delta\theta_L   + G \cos\phi \delta\theta_R - G \cos\phi \delta\theta_L  + \alpha G \cos\phi (\delta\varphi_L - \delta\varphi_R);\\
   &(1 + \alpha^2)\frac{d \delta \theta_R}{d \tilde{t}} =   G \cos \phi  (\delta\varphi_R - \delta\varphi_L) - \alpha \left[G \cos\phi \left[\delta \theta_L - \delta\theta_R \right] + \tilde{K} \delta\theta_R \right],\\
   & (1 + \alpha^2)  \frac{d \delta \varphi_R}{d \tilde{t}} = \tilde{K} \delta\theta_R   + G \cos\phi \delta\theta_L - G \cos\phi \delta\theta_R  + \alpha G \cos\phi (\delta\varphi_R - \delta\varphi_L) . \label{eq:LLG_Angles_dP}
\end{align}
The eigenvalues of this linear system up to linear order in $\alpha$ read as:
\begin{align}
    &\lambda_1 = 0,\qquad \lambda_2 = - \frac{\alpha \tilde{K}}{1 + \alpha^2} \approx - \alpha \tilde{K},\\
    &\lambda_{3,4} = \frac{1}{2 (1 + \alpha^2)} \left\{ - \alpha \left( \tilde{K} - 4 G \cos \phi\right) \mp \sqrt{\alpha ^2 \tilde{K}^2 +8 G \cos \phi \left( \tilde{K}  - 2 G \cos \phi \right) } \right\} \approx \nonumber \\
    &\qquad\qquad\qquad \approx \frac{1}{2 } \left\{ - \alpha \left( \tilde{K} - 4 G \cos \phi\right) \mp  \alpha \tilde{K} \right\} . \label{eq:Lambdas_b}
\end{align}
We use that, for realistic system parameters like those discussed in the main text, $\alpha^2 \tilde{K} \gg 8G$ . This inequality holds because magnetic anisotropy field greatly exceeds the Josephson torque, and even the smallness of $\alpha^2$ cannot bring them to comparable scales. Consequently, the stability of the parallel configuration is governed by $\lambda_4 = 2G\alpha \cos\phi$, which implies that the P state is stable for $\cos\phi < 0$ and unstable otherwise. Moreover, if one repeats the procedure given by Eqs.~(\ref{eq:LLG_Angles_aP})--(\ref{eq:Lambdas_b}) for the antiparallel FI configuration, then the corresponding eigenvalue which determines the stability of the AP configuration is $\lambda = - \frac{2 G}{\alpha} \cos\phi$. We again conclude that when $\cos \phi > 0$, then the AP state is stable and when $\cos \phi < 0$,  it is not stable. 


\end{document}